\documentclass[runningheads]{llncs}
\usepackage[linesnumbered,ruled, vlined]{algorithm2e}
\usepackage{graphicx}
\usepackage{mathtools}
\usepackage{hyperref}
\usepackage{pgfplots}
\usepackage[linesnumbered,ruled, vlined]{algorithm2e}
\usepackage{setspace}
\usepackage{hyperref}
\usepackage{verbatim}
\usepackage{caption}
\usepackage{subcaption}
\pgfplotsset{compat=1.9}
\usepgfplotslibrary{groupplots}
\usetikzlibrary{matrix}
\usepackage{xcolor}

\SetCommentSty{mycommfont}
\begin{document}
\title{Dimensional Data KNN-based Imputation\thanks{This work is supported by the French National Research Agency (ANR), Project ANR-19-CE23-0005 BI4people (Business intelligence for the people).}}
%
%
\author{Yuzhao Yang\inst{1}\orcidID{0000-0002-6552-4812} \and
Jérôme Darmont \inst{2}\orcidID{0000-0003-1491-384X}\and
Franck Ravat\inst{1}\orcidID{0000-0003-4820-841X}\and  Olivier Teste\inst{1}\orcidID{0000-0003-0338-9886}}
\authorrunning{Y. Yang et al.}
\institute{IRIT-CNRS (UMR 5505), 
    Université de Toulouse, France \\\email{\{Yuzhao.Yang, Franck.Ravat, Olivier.Teste\}@irit.fr} \and
Université de Lyon, Lyon 2, UR ERIC, France \\
\email{jerome.darmont@univ-lyon2.fr}}

%

%
\maketitle              
\begin{abstract}
Data Warehouses (DWs) are core components of Business Intelligence (BI). Missing data 
in DWs 
have a great impact on data analyses. Therefore, missing data need to be completed. Unlike other existing data imputation methods mainly adapted for facts, we propose a new imputation method for dimensions. This method contains two steps: 1) a hierarchical imputation and 2) a k-nearest neighbors (KNN) based imputation. Our solution has the advantage of taking into account the DW structure and dependency constraints. Experimental assessments validate our method in terms of effectiveness and efficiency.


\end{abstract}
\section{Introduction}

Data warehouses (DWs) are widely used in companies and organizations as a significant Business Intelligence (BI) tool to help them building their decision support systems. Data in DWs are usually modelled in a multidimensional way, which allows the user to analyse data through On Line Analytical Processing (OLAP). An OLAP model organizes data according to analysis subjects (facts) associated to analysis axis (dimensions). Each fact is composed of measures. Each dimension contains one or several analysis viewpoints (hierarchies). 

Missing data may exist in a DW. There are 2 types of DW missing data: \textbf{dimensional missing data} which are missing data in the dimensions and \textbf{factual missing data} which are in the facts. These missing data have impact on OLAP analyses. It is important to complete the missing data for the sake of a better data analysis.



Data imputation is the process 
of replacing the missing values by some plausible values based on information available in the data \cite{Li04}. 
The current DW data imputation research mainly focuses on factual data \cite{wu02,Ribeiro11,bimont20}. Yet the dimensional missing data 
make aggregated data incomplete and make it hard to analyse them with respect to hierarchy levels. Therefore the imputation for DW dimensions is also necessary. However the DW dimension has a complex structure containing different hierarchies with different granularity levels having their dependency relationships. When we complete the dimensional missing data, we have to take the DW structure and the dependency constraints into account. We proposed a hierarchical imputation based on the inter- and intra-dimensional hierarchical dependency relationships \cite{yang21} for the imputation of dimensional missing data. To the best of our knowledge, there is no other specific data imputation method for DW dimensions. The hierarchical imputation is convincible because we use accurate data based on real functional dependency relationships. However, this method is limited owing to the sparsity problem which means that for an instance to be completed, there may not be an instance sharing the same value on a lower-granularity level of the hierarchy.

In order to complete as many values as possible, in this paper, we propose H-OLAPKNN, an imputation method for DW dimensions by extending the hierarchical imputation with a novel dimension imputation method called OLAPKNN. OLAPKNN is based on K-nearest neighbours (KNN) algorithm. KNN imputation finds the K nearest neighbors of an instance with missing data then fills in the missing data based on the mean or mode of the neighbors' value \cite{Troyanskaya01}. We choose KNN because it is a non-parametric and instance-based algorithm, which is widely applied for data imputation \cite{beretta2016nearest} and has been proved to have relatively high accuracy 
\cite{LI201464,Troyanskaya01}. Compared to the basic KNN imputation, OLAPKNN considers the structure complexity and the dependency constraints of the dimension hierarchies. Moreover, the dimensional data are usually qualitative on which we focus in this paper.




The remainder of this paper is organized as follows. In Section~\ref{sec:relatedworks}, we review the related work about data imputation algorithms. In Section~\ref{sec:prel}, we formalize the DW dimension model. In Section~\ref{sec:distance}, we propose a distance calculation method for dimension instances. In Section \ref{sec:imputation}, we explain in detail our proposed 
dimension imputation algorithm. In Section~\ref{sec:expe}, we validate our proposal by some experiments. 
In Section~\ref{sec:conclu}, we conclude this paper and hint at future research.




\section{Related Work}
\label{sec:relatedwork}

There are various data imputation methods \cite{Miao18}: statistic based imputation, machine-learning based imputation, rule based imputation, external source based imputation and hybrid methods etc. The statistic based imputation completes the missing values by applying the statistical methods like filling average, the most frequent value or with the value of the most similar record; there are also methods using the regression to predict the missing values \cite{Gar10}. The machine learning based imputation methods use algorithms like k-nearest neighbor (KNN) \cite{LI201464,Troyanskaya01,GARCIALAENCINA20091483,pan2015}, regression models \cite{little2002statistical}, Naive Bayes \cite{Far07} to predict the missing values. The rule based imputation methods \cite{Fan10,Son15,breve2022renuver} complete the missing values by some business rules, similarity rules or dependency rules. Concerning the external source based methods, 
the crowdsourcing \cite{Lod13} can be applied for the data imputation by putting forward the queries in the crowdsourcing frameworks and collecting answers to complete the missing data. There are also methods which realize the imputation through web information \cite{Li14,Yak12} like web pages, web lists and web tables. What's more, there are hybrid methods which mix different imputation methods to provide a higher performance.

The statistic and machine learning based methods mainly focus on the numerical data, which fit for the imputation of facts where the data are mostly numerical. However, in the dimensions, there are mainly qualitative data which make it difficult to process the data imputation by such imputation methods. The rule based and external source based imputation methods may be suitable for the imputation of dimensions, but they need time and efforts to create rules or find the appropriate sources. Hence we propose H-OLAPKNN which combines the hierarchical imputation with a KNN-based imputation method.

\label{sec:relatedworks}
\section{DW Dimension}
\label{sec:prel}

As a DW is composed of dimensions and facts and we focus on the dimension imputation, we introduce the DW dimension concepts used in this paper~\cite{Ravat08}. 


\begin{definition}[Dimension]
In a data warehouse, a \textbf{dimension}, denoted by $D$, is defined as ($A^D, H^D,$ $I^D$). $A^D = \{ a_1,...,a_u\} \cup \{id\}$ is a set of attributes, where $id$ represents the dimension's identifier; $H^D = \{H_1,...,H_v\}$ is a set of hierarchies; $I^D$ is a matrix of dimension instances, for a given row $r$, the row instance vector is denoted as $i_r$; for a given attribute $a_u$, their joint instance value is denoted as $i_{r, a_u}$.

\end{definition}

\begin{definition}[Hierarchy]
A \textbf{hierarchy} of dimension $D$, denoted by $H \in H^D$, is defined as $(Param^{H}, Weak^{H})$. $Param^{H} = <id^D, p^H_2, ..., p^H_v$ $>$ is an ordered set of dimension attributes, called \textbf{parameters}, which set granularity levels along the dimensions, $\forall k \in [1...v], p^H_k \in A^D$. 
Parameter $p^H_1$ rolls up to $p^H_2$ in $H$ is denoted as $p^H_1 \preceq_H p^H_2$; $Weak^{H}$ = $Param^{H} \rightarrow 2^{(A^D - Param^{H})}$ is a mapping possibly associating each parameter with one or several \textbf{weak attributes}, which are also dimension attributes providing additional information; All parameters and weak attributes of $H$ constitute the hierarchy attributes of $H$, denoted by $A^H = Param^H \cup (\bigcup\limits_{p_v^H \in Param^H} Weak^H[p_v^H])$ 
\end{definition}

There exists different types of hierarchy, but the most basic and common one is the strict hierarchy \cite{Malinowski04} where a value at a hierarchy's lower-granularity belongs to only one higher-granularity value \cite{trujillo2001}. Thus in this paper, we only consider the case of the strict hierarchy.

\section{Distance Between Dimension Instances}
\label{sec:distance}
Since the KNN imputation select the $k$-nearest neighbors of the missing data instance for the imputation, we should calculate the distance between dimension instances containing missing data to be completed and other instances. In a dimension $D$, for an instance $i_1 \in I^D$ containing missing data on a hierarchy $H_1 \in H^D$, and another instance $i_2 \in I^D$, we propose to calculate their distance by 4 levels:
\begin{itemize}

    \item The \textbf{dimension instance distance} is the final distance between two instances $i_1$ and $i_2$, denoted by $\Delta(i_1, i_2)$. Since the attributes on the same hierarchy have their dependency relationships, we consider the attributes of a hierarchy as an entirety. $\Delta(i_1, i_2)$ is thus calculated by the weighted sum of the \textbf{hierarchy instance distances}. 
    
    \item The \textbf{hierarchy instance distance} is the distance of the attributes of a hierarchy $H_2 \in H^D$ i.e. distance between $\{i_{1,a_1} \in i_1 : a_1 \in A^{H_2}\}$ and $\{i_{2,a_1} \in i_2 : a_1 \in A^{H_2}\}$, denoted by $\Delta_{H_2}{(i_1, i_2)}$. It is calculated by the weighted sum of the \textbf{hierarchy level instance distances}. The lowest-granularity level of each hierarchy is the same i.e. $id$ with its weak attributes, so we consider the hierarchy instance distance from the second level of the hierarchy and we regard each weak attribute of $id$ as a hierarchy containing only one parameter.
    
    \item The \textbf{hierarchy level instance distance} is the instance distance between the attributes of a level $l$ on a hierarchy $H_2$ i.e. distance between $\{i_{1,a_2} \in i_1 : a_2 \in p_l^{H_2} \cup Weak^{H_2}[p_l^{H_2}] \}$ and $\{i_{2,a_2} \in i_2 : a_2 \in p_l^{H_2} \cup Weak^{H_2}[p_l^{H_2}]\}$, denoted by $\Delta_{p_l^{H_2}}{(i_1, i_2)}$. It is calculated by the average of the instance distances of the level's parameter and weak attributes (\textbf{attribute distances}).
    
    \item The \textbf{attribute distance} is the instance distance of an individual attribute $a_u \in A^D$ i.e. distance between $i_{1,a_u}$ and $i_{2,a_u}$, denoted by $\Delta(i_{1,a_u}, i_{2,a_u})$.
    
\end{itemize}


Based on the explanation of the distances, we then give the formulas and some examples to illustrate them in detail.

\begin{figure}
\centering
\begin{subfigure}{0.2\textwidth}
  \centering
  \includegraphics[width=1.3\linewidth]{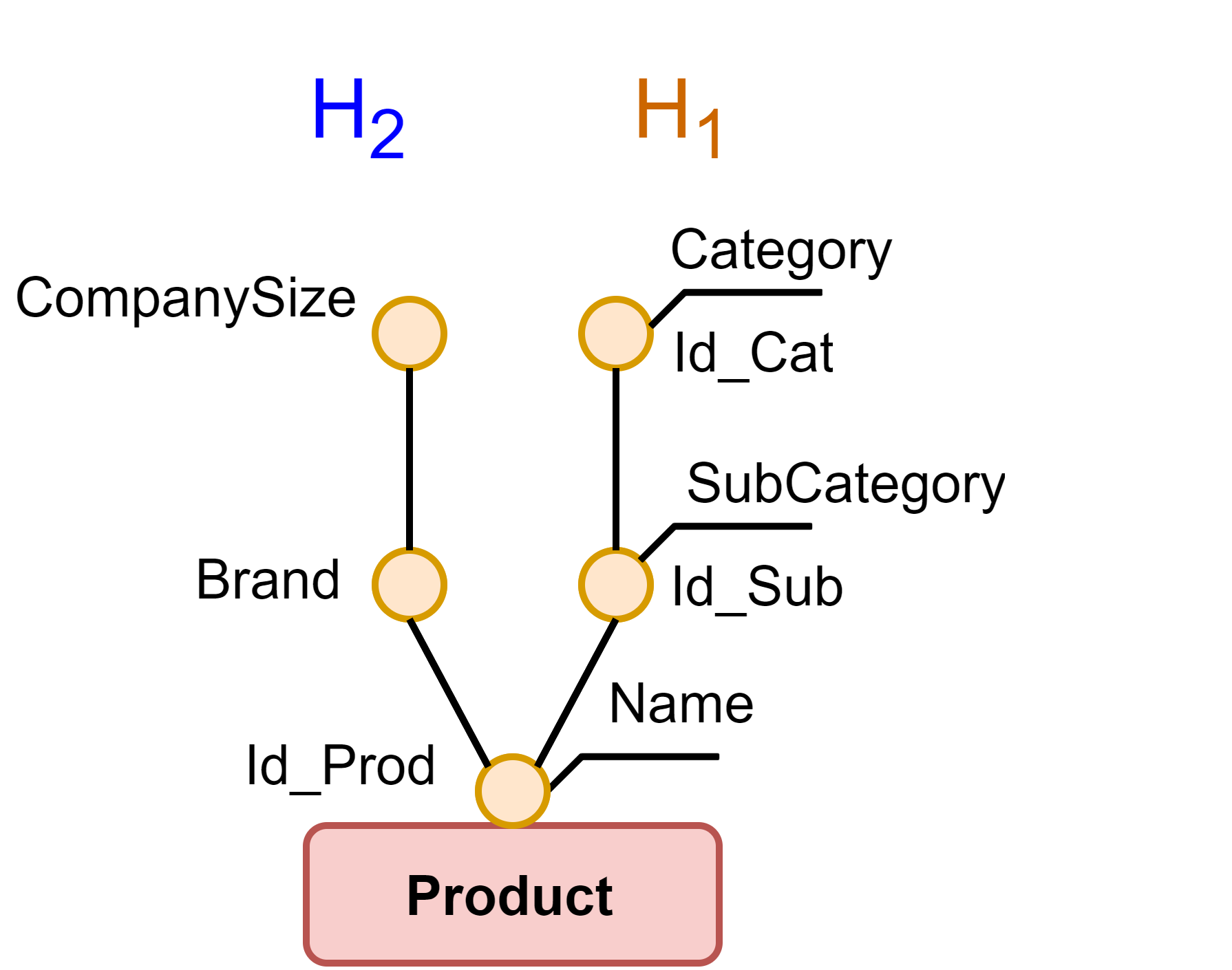}
  \caption{Schema}
  \label{fig:sub1}
\end{subfigure}%
\begin{subfigure}{.8\textwidth}
  \centering
  \includegraphics[width=\linewidth]{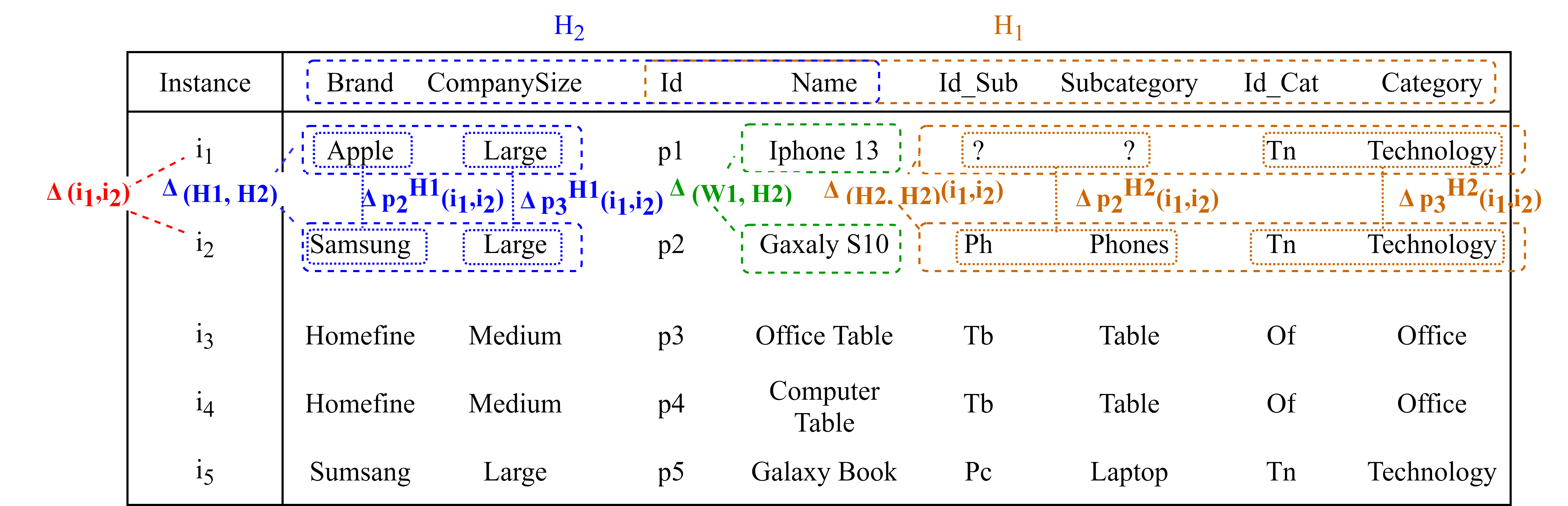}
  \caption{Instances}
  \label{fig:sub2}
\end{subfigure}
\caption{Schema and instances of dimension $Product$}
\label{fig:schemainstance}
\end{figure}

\begin{figure}[h]
\centering
 \includegraphics[width=0.9\linewidth]{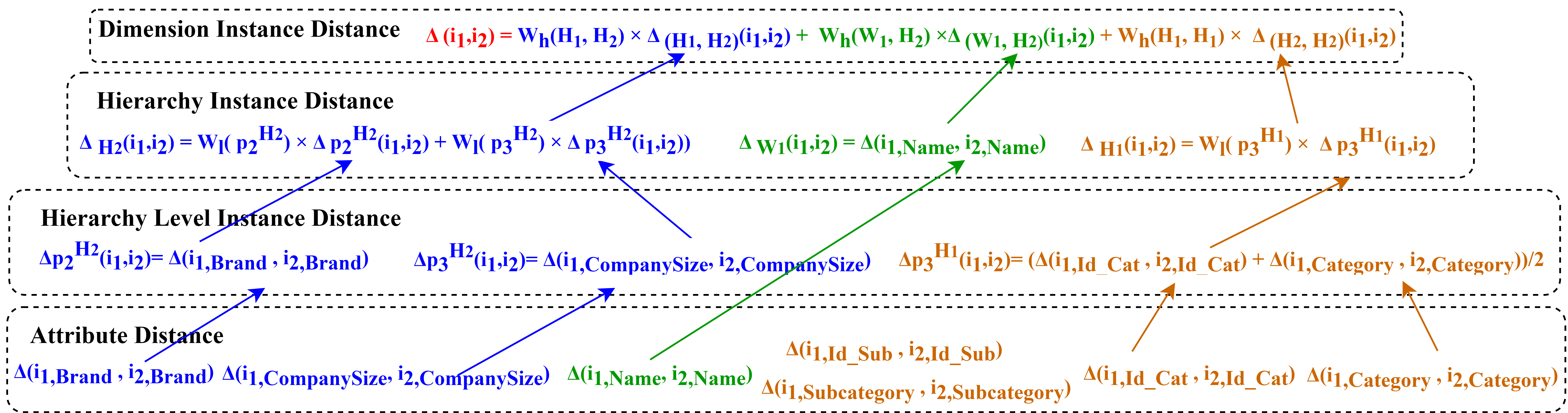}
 \caption{Distance between $i_1$ and $i_2$}  
 \label{fig:distance}
\end{figure}

\begin{example}
Given a dimension $Product$ containing two hierarchies $H_1$ and $H_2$ whose schema and instances are shown in Fig. \ref{fig:schemainstance}. Instance $i_1$ contains missing values on $H_1$, Fig. \ref{fig:distance} shows the calculation of the distance $\Delta(i_1, i_2)$ between $i_1$ and another instance $i_2$.
\end{example}

\subsection{Attribute Distance}
\label{sec:distancemetric}

There are different attribute data types which can be mainly classified into numerical and textual. For numerical data, we use normalized distance of numerical data \cite{Han2012}. For textual data, we first apply semantic distance e.g. cosine distance based on word2vec \cite{JATNIKA2019160}. If the attribute value cannot be found in the model, we can then use the syntactic distance e.g. normalized Levenshtein Distance \cite{Yujian07}.

For an attribute $a_{u1}$, if $i_{1, a_{u1}}$ is missing, then $\Delta(i_{1,a_{u1}}, i_{2,a_{u1}})$ cannot be calculated and is not taken into count for the distance calculation. For an attribute $a_{u2}$, if $i_{2, a_{u2}}$ is missing, then $\Delta(i_{1,a_{u2}}, i_{2,a_{u2}})$ is obtained by the average distance between $i_{1,a_{u2}}$ and other instances whose value of $a_{u2}$ is not missing.

\begin{example}
Following the calculation rules of the attribute distance, we obtain $\Delta(i_{1,Brand}, i_{2,Brand}) = 0.71$, $\Delta(i_{1,CompanySize}$, $i_{2,CompanySize}) = 0$, $\Delta(i_{1,Name},$ $i_{2,Name}) = 0.8$, $\Delta(i_{1,Id\_Cat}, i_{2,Id\_Cat}) = 0$, $ $ $\Delta(i_{1,Category}, i_{2,Category}) = 0$. Since $i_{1,Id\_Sub}$ and $i_{1,Subcategory}$ are missing, $  $ $\Delta(i_{1,Id\_Sub}, i_{2,Id\_Sub})$ and $\Delta(i_{1,Subcategory},$ $i_{2,Subcategory})$ cannot be calculated and are not taken into count for the calculation of $\Delta(i_1, i_2)$.
\end{example}

\subsection{Hierarchy Level Instance Distance}

The hierarchy level instance distance $\Delta_{p_l^{H_2}}{(i_1, i_2)}$ is calculated as (\ref{leveldistance}).

\begin{equation}\label{leveldistance}
\Delta_{p_l^{H_2}}{(i_1, i_2)}  = \frac{\Delta(i_{1,p_l^{H_2}}, i_{2, p_l^{H_2}}) + \sum\limits_{w \in Weak[p_l^{H_2}]}\Delta(i_{1,w}, i_{2, w})}{1 + |Weak[p_l^{H_2}]|}
\end{equation}

As we mentioned that we only consider the levels from the second level of each hierarchy, we do not calculate the distance for the first level of hierarchies. The weak attributes of the first hierarchy levels are regarded as hierarchies containing only one parameter, so their level distance is not needed to be calculated neither.

\begin{example}
According to (\ref{leveldistance}), for the levels in $H_1$, we have $\Delta^{H_1}_{p_3}(i_1,i_2) = (0 + 0) / 2 = 0$. As the parameter and weak attribute value of the second level $i_{1,Id\_Sub}$ and $i_{1,Subcategory}$ are missing, the distance of this level is not taken into account. For $H_2$, since the two levels contain only one parameter without weak attribute, their hierarchy level is equal to the attribute distance of the parameter, so we have $\Delta^{H_2}_{p_2}(i_1,i_2) = 0.71$, $\Delta^{H_2}_{p_3}(i_1,i_2) = 0$.
\end{example}

\subsection{Hierarchy Instance Distance}

The hierarchy instance distance is calculated as (\ref{hierarchydistance}), where $W_l(p_l^{H_2})$ is the hierarchy level weight.

\begin{equation}\label{hierarchydistance}
\Delta_{H_2}(i_1, i_2) = \sum\limits_{p_l^{H_2} \in H_2 \setminus \{id\}} W_l(p_l^{H_2}) \Delta_{p_l^{H_2}}(i_1, i_2)
\end{equation}

For a weak attribute $w \in Weak^{H_2}[id]$ of the first hierarchy level, $\Delta_{w}(i1, i_2) = \Delta(i_{1,w}, i_{2,w})$.

\subsubsection{Hierarchy Level Weight}
Since the parameters on the lower levels have thinner granularity, their weight for measuring the hierarchy instance distance should be higher. Here, we propose two hierarchy level weights: one is based on the cardinalities of the parameters and another is an incremental weight. 

\begin{itemize}
    \item For the cardianlity-based weight, we consider the number of the distinct values of the level as the portion of the weight. Thus for the cardianlity-based hierarchy level weight of the $l$th level at $H_2$ is calculated as (\ref{wc}), where $dv(n)$ denotes the number of distinct values of the $n$th level.
    \begin{equation}\label{wc}
	W^c_l(p_l^{H_2}) = \frac{dv(l)}{\sum_{j=2}^{|Param^{H_2}|} dv(j)}
    \end{equation} 
    \item However, when the cardinality ratio between certain parameters is very large, the cardinality-based weight may be biased. So we also propose another type of incremental hierarchy level distance weight. For the incremental weight, we consider the weight of the highest-granularity as one portion and it increases by one portion for each neighboring lower-granularity level. The total weight should be equal to 1, thus the incremental hierarchy level weight of the $l$th level at $H_2$ is calculated as (\ref{wi}).

    \begin{equation}\label{wi}
    	W^i_l(p_l^{H_2}) = \frac{2(|Param^{H_2}|-l+1)}{|Param^{H_2}|^2-|Param^{H_2}|}
    \end{equation} 
\end{itemize}

\begin{example}
Our example has only 5 instances, so we can use cardinality-based weight to get hierachy level weight. We thus have for $H_1$: $W_l(p^{H_1}_2) = 3 / (3 + 2) = 0.6$ and $W_l(p^{H_1}_3) = 2 / (3 + 2) = 0.4$. For $H_2$:  $W_l(p^{H_2}_2) = 3 / (3 + 2) = 0.6$ and $W_l(p^{H_2}_3) = 2 / (3 + 2) = 0.4$. We can then calculate the hierarchy instance distances: $\Delta_{H_1}(i_1, i_2) = 0.4 \times 0 = 0$, $\Delta_{H_2}(i_1, i_2) = 0.6 \times 0.71 + 0.4 \times 0 = 0.426$, $\Delta_{w_1}(i_1, i_2) = 0.8$.

\end{example}

\subsection{Dimension Instance Weight}

The dimension instance weight $\Delta(i_1, i_2)$ is calculated as (\ref{dimensiondistance}), where $W_h(H_1, H_2)$ and $W_h(H_1, w)$ are hierarchy weights of $H_2$ and $w$ with respect to $H_1$.

\begin{equation}\label{dimensiondistance}
\Delta(i_1, i_2) = \sum\limits_{H_2 \in H^D} W_h(H_1, H_2) \Delta_{H_2}(i_1, i_2) + \sum\limits_{w \in Weak^{H_2}[id]} W_h(H_1, w) \Delta_{w}(i_1, i_2)
\end{equation}

\subsubsection{Hierarchy Weight}

The dependency degree in the rough set theory \cite{PAWLAK20073} 
measures the degree of the dependency between attributes, so it is applied for the hierarchy weight.

When calculating the hierarchy distance weight, we can consider a decision system $S = (I^D,A^{H_2}_{n},A^{H_1}_{n})$, since we do not take the first level of a hierarchy into account, $A^{H_1}_{n} = A^{H_1} \setminus (\{id\} \cup Weak^{H_1}[id])$, $A^{H_2}_{n} = A^{H_2} \setminus (\{id\} \cup Weak^{H_2}[id])$. The second hierarchy level parameters $p_2^{H_1}, p_2^{H_2}$ determine all the other hierarchy attributes in $A^{H_1}_{n}$ and $A^{H_2}_{n}$, we can reduce the attribute sets of $A^{H_1}_{n}$ and $A^{H_2}_{n}$ to the sets containing only the values of the second hierarchy level parameter $p_2^{H_1}, p_2^{H_2}$.
According to \cite{PAWLAK20073}, the degree $k$ to which $H_1$ depends on $H_2$, denoted $H_2 \Rightarrow_k H_1$ is thus defined as: 
\begin{equation}\label{DD}
	k = \gamma (A^{H_2}_{n},A^{H_1}_{n}) = \gamma (p_2^{H_2},p_2^{H_1}) =  \frac{card(POS_{p_2^{H_2}}(p_2^{H_1}))}{card(I^D)}
\end{equation}
where 
$POS_{p_2^{H_2}}(p_2^{H_1}) = \bigcup_{X \in I^D / p_2^{H_1}}p_{2*}^{H_2}(X) $
and $card(X)$ is 
the cardinality of an non-empty set $X$, the missing second level parameter values are not taken into account. For $H_1$ itself, we have $\gamma (A^{H_1}_{n},A^{H_1}_{n})=1$.


The hierarchy distance weight of $H_2$ with respect to $H_1$ is the ratio of their dependency degree with respect to the sum of the dependency degrees of the all hierarchies and first level weak attributes in $D$ with respect to $H_1$ as (\ref{Wh}).
\begin{equation}\label{Wh}
W_h(H_1, H_2) = \frac{\gamma (A^{H_2}_{n},A^{H_1}_{n})}{\sum\limits_{H_3 \in H^D}\gamma (A^{H_3}_{n},A^{H_1}_{n}) + \sum\limits_{w \in Weak^{H_1}[id]}\gamma (w,A^{H_1}_{n})}
\end{equation}

\begin{example}
In our example, we have $card(I^D) = 5$, $card(POS_{p_2^{H_2}}(p_2^{H_1}))  = 2 $, so $\gamma (A^{H_2}_{n},A^{H_1}_{n}) = 2/5 = 0.4$. In the same way, we can get $\gamma (w_1,A^{H_1}_{n}) = 2/5 = 0.4$, we also have  $\gamma (A^{H_1}_{n},A^{H_1}_{n}) = 1$. We can thus get the hierarchy weights: $W_h(H_1,H_2) = 0.4/(0.4 + 0.4 +1) = 0.22$, $W_h(H_1,H_1) = 1/(0.4 + 0.4 +1) = 0.56$ and $W_h(w_1,H_2) = 0.4/(0.4 + 0.4 +1) = 0.22$. We can finally obtain the dimension instance distance $\Delta(i_1, i_2) = 0.22 \times 0.46 + 0.22 \times 0.8 + 0.56 \times 0 = 0.28$
\end{example}


\section{H-OLAPKNN Imputation}
\label{sec:imputation}
\subsection{H-OLAPKNN Overview}




The H-OLAPKNN imputation is shown in Algo. \ref{algo:H-OLAPKNN}. It is composed of three steps where the first is the hierarchical imputation and the next two steps concern the OLAPKNN imputation.

\begin{enumerate}
    \item 
    The hierarchical imputation is based on the functional dependencies of the hierarchy attributes. It searches for an instance having the same value on a lower-granularity level parameter of the missing value and whose attribute of the missing value is not empty, we can then replace the missing value with this non-empty value ($line_1$). 
    
    \item The weak attributes' values are determined by their parameters' values, so we complete the parameters before completing their weak attributes. Thus then, for missing data of each hierarchy ($line_2$), we create candidate lists of the instances containing possible replaced values and select the $k$ nearest neighbors in the candidate lists to complete the missing data ($line_3$).
    \item There are weak attributes which can be completed together with their parameter. Finally for the remaining missing weak attribute data, they are completed in the similar way ($line_4$).
\end{enumerate}

Next, we explain in detail the OLAPKNN imputation algorithm. A weak attribute of a hierarchy can be regarded as a ``highest level parameter'' of a part of the hierarchy whose imputation is similar to the parameter imputation. So we only explain the parameter imputation in this paper.

\subsection{Imputation for Parameters by OLAPKNN}
\subsubsection{Parameter Imputation Order}
We first introduce the continuous missing parameter group in order to explain the imputation order for parameters.
 \begin{definition}[Continuous missing parameter group]
For an instance $i_r \in I^D$ in the dimension $D$ containing missing values on parameters of a hierarchy $H$, all these parameters are in a set $Pm^H_r = \{p_{v}^H \in Param^H : i_{r,p_{v}^H}$ is empty$\}$. For the parameters in $Pm^H_r$, they can be divided into one or several continuous missing parameter groups. A \textbf{continuous missing parameter group (CG)} contains one or several parameters which are neighbors on $H$ and are maximal neighbors in $Pm^H_r$. By neighbors on $H$, we mean that for the parameter $p_{lowest}$ having the lowest-granularity level in the CG on $H$ and the one $p_{highest}$ having the highest-granularity level, if there exists any parameter $p_{middle} \in Param^H$, such that $p_{lowest} \preceq_H p_{middle} \preceq_H p_{highest}$, then $p_{middle} \in Pm^H_r$; By maximal neighbors in $Pm^H_r$, we mean that if there exists any parameter $p_{low_2} \in Param^H$, such that $p_{low_2} \preceq_H p_{lowest}$, then $p_{low_2} \not\in Pm^H_r$, if there exists any parameter $p_{high_2} \in Param^H$, such that $p_{highest} \preceq_H p_{high_2}$, then $p_{high_2} \not\in Pm^H_r$. We call all CGs of a hierarchy $H$ containing a same number of parameters a n-CGs of $H$, where $n$ denotes the number of parameters.
\label{def:continuous}
\end{definition}

Algo. \ref{algo:imputeParam} shows the imputation of the parameters. For a given hierarchy $H$ on a dimension $D$, we carry out the imputation for parameters in the $n$-CGs by the ascending order of $n$ $(line_1)$. We can thus make sure that all the $(n-1)$-CGs instances are completed so that we can carry out the imputation for the $n$-CGs based on the existing data. 
Then for each $n$-CGs, we look at all possible CG combinations $(line_{2 - 3})$. Next we verify if there are instances containing missing values for each possible CG $(line_{4 - 9})$. According to $Definition$ \ref{def:continuous}, the instances of a CG on $H$ have missing values on all parameters of the group. If there is a neighboring lower-granularity or higher-granularity parameter of the group, the instances do not have missing value on them $(line_9)$.

\noindent
\scalebox{0.65}{
\begin{minipage}{0.75\linewidth}
\begin{algorithm}[H]
$hierarchicalImputation (D) $\;


\For{$H \in H^D$}
    {
        $imputeParam(D, H)$ \;
        $imputeWeak(D, H)$ \; 
    }
   
\caption{$H-OLAPKNN(D)$}
\label{algo:H-OLAPKNN}
\end{algorithm}
\end{minipage}
}
\scalebox{0.65}{
\begin{minipage}{0.75\linewidth}
\begin{algorithm}[H]
\For{$ncontinuous \gets 1$ to $|Param^H| - 1$}
    {
    \For{$i \gets 1$ to $|Param^H| - ncontinuous$}
        {
        $P_{CG} \gets Param^H[i:i+ncontinuous-1]$ \;
        $p_{low}, p_{high} \gets \o$ \;
        \If{$i > 1$}
            {
            $p_{low} \gets Param[i - 1]$ \;
            }
        \If{$i < |Param^H| - ncontinuous$}
            {
            $p_{high} \gets Param[i + ncontinuous]$ \;
            }
        $I_{missing} = \{i_r \in I^D:(\forall p_{cg} \in P_{CG}, i_{r,p_{cg}} = null) \land (\exists p_{low} \implies i_{r,p_{low}} \neq null) \land (\exists p_{high} \implies i_{r,p_{high}} \neq null) \}$ \;
        $lowMap \gets Map$ \;
        \For{$i_m \in I_{missing}$}
            {
            $I_{candidate} \gets getCandidateList( D, P_{CG},$ $p_{high}, i_m, 1)$ \;
            
            
            $vWeightMap \gets getVWeightMap(D, i_m,$ $I_{candidate}, k, P_{CG})$ \;
            
            $lowMap \gets replaceNoPlow(D, H,$ $lowMap, vWeightMap, i_m, P_{CG}, p_{low})$ \;
                
            }
        \If{$\exists p_{low}$}
            {
            $replacePlow(lowMap, P_{CG}, H, D, p_{low})$ \;
            }
        }
    }
\caption{
$imputeParam(D, H)$}
\label{algo:imputeParam}
\end{algorithm}
\end{minipage}
}

\subsubsection{Candidate List}

Since some missing data are already completed by the hierarchical imputation, for the remaining missing data, they can no longer be completed with the aid of their lower-granularity parameters. 
For a value of one parent parameter, there may be several possible values on a child parameter of its. So for a missing data instance of a CG, we can find all possible replaced values based on their neighboring higher parameter 
and create a candidate list (Algo. \ref{algo:imputeParam} $line_{11}$). The candidate list contains not only the candidate replaced values of CG attributes but also the values of all other attributes of the dimension because we need all attribute's value for the calculation of the distances.

Algo. \ref{algo:getCanList} shows the candidate list creation for an instance of a CG. If the neighboring higher-granularity parameter $p_{high}$ of the CG exists, we search for all the instances having the same values on $p_{high}$ as the CG instance, and containing non-missing values on the CG parameters. Then these instances can be added into the candidate list ($line_{1-3}$). If there does not exist a neighboring higher parameter for a CG, we add all the instances of the dimension which contain non-missing values on the CG parameters into the candidate list ($line_{4-5}$). 



\subsubsection{Creation of Replaced Value Weight Map}

For the CG instance, we can get a map for each possible replaced values in the nearest neighbors with their distance-based weight for the selection of the final replaced value as described in Algo. \ref{algo:getVWeightMap}.
We first create a map of each instance in the candidate list with its distance with respect to the missing instance ($line_{1-3}$). Then we can select the $k$ nearest candidate instances to create a candidate list if the candidate list contains more than $k$ instances, if not, we can keep all candidate instances ($line_{4-5}$). The selected candidate instances may contain same replaced values, so we create a map of each replaced values with their weight ($line_6$). According to \cite{dudani1976}, for an instance $i_m$ of a CG, for a selected candidate list containing $k$ instances, the distance weight of the $n$ nearest instance $i_{cn}$ can be calculated as (\ref{WeightV}), where $i_{ck}$ denotes the $k$th nearest instance and $i_{c1}$ denotes the nearest instance. It is to be noted that $W_d (i_m, i_c) = 1$ when $\Delta(i_m, i_{ck}) = \Delta(i_m, i_{c1})$.

\begin{equation}\label{WeightV}
W_d (i_m, i_c) = \frac{\Delta(i_m, i_{ck}) - \Delta(i_m, i_{cn})}{\Delta(i_m, i_{ck}) - \Delta(i_m, i_{c1})}
\end{equation}

Thus the weight of a candidate of replaced values is the sum of the weight of the instances which contain them ($line_{4-5}$).

\noindent
\scalebox{0.65}{
\begin{minipage}{0.75\linewidth}
\begin{algorithm}[H]
\If{parameter = 1}
    {
        \If{$\exists p_{high}$}
            {
            $I_{candidate} \gets \{i_r \in I^D:(\exists p_{cg} \in P_{CG}, i_{r,p_{cg}} \neq null) \land (i_{r,p_{high}} = i_{r_{missing},p_{high}}) \}$ \;
            }
        \Else
            {
            $I_{candidate} \gets \{i_r \in I^D :(\exists p_{cg} \in P_{CG}, i_{r,p_{cg}} \neq null) \}$ \;
            }
    }
\Else{
$I_{candidate} \gets \{i_r \in I^D:( i_{r,weak} \neq null) \}$ \;
}

\Return{$I_{candidate}$}
\caption{$getCanList(D, P_{CG}, p_{high}, i_m, parameter)$}
\label{algo:getCanList}
\end{algorithm}

\end{minipage}
}
\scalebox{0.65}{
\begin{minipage}{0.75\linewidth}
\begin{algorithm}[H]
$iDistanceMap \gets Map$ \;

\For{$i_c \in I_{candidate}$}
    {
    $iDistanceMap[i_{c, id}] \gets \Delta(i_m, i_c)$ \;
    }
\If{$|I_{candidate}| > k$}
    {
    $iDistanceMap \gets iDistanceMap.top(k)$\;
    }
$vWeightMap \gets Map$ \;
\For{$i_{c, id} \in iDistanceMap.keys()$}
    {
    \tcc{addMap(Map, key, value): Create the map if it does not exist. Add the value to the existing value if the key exists, assign the value to the key if not.}
    $addMap(vWeightMap, \{i_{c, p_{cg}}:i_{c, p_{cg}} \in i_{c, P_{CG}}\}, Wv(i_m, i_c))$\;  
    }
\Return{vWeightMap}
\caption{$getVWeightMap(D, i_m,I_{candidate}, k, P_{CG})$}
\label{algo:getVWeightMap}
\end{algorithm}
\end{minipage}
}

\subsubsection{Replacement of Values}

To fill in the missing values of CG, we have two cases: the first case (Algo. \ref{algo:imputeParam} $line_{13}$) is that there is no lower non-id parameter of the missing parameter group, the second case (Algo. \ref{algo:imputeParam} $line_{14-15}$) is that there is such parameter. The difference is that for the second case, we have to take the strictness of hierarchy into account by making sure that a lower parameter value of the CG has only one higher-granularity level parameter after the imputation.

The replacement of the values of the first case is described in Algo. \ref{algo:replaceNoPlow}. We can take the values having the highest weight in the weight map ($line_1$) to fill in the missing values of the CG ($line_{2-3}$).

The replacement of the values for the second case is described in Algo. \ref{algo:replaceNoPlow} and Algo. \ref{algo:replacePlow}. We create a map $lowMap$ for each neighboring lower-granularity parameter value which corresponds to another map of the each possible replaced value and its total weight (Algo. \ref{algo:imputeParam} $line_{10}$). For each instance of the CG, we get the replaced values with the highest value weight (Algo. \ref{algo:replaceNoPlow} $line_1$). For the value of its neighboring lower-granularity parameter, we update the replaced values and the weight (Algo. \ref{algo:replaceNoPlow} $line_{8-10}$). When all the missing instances of a CG are treated, we get a final $lowMap$. For each value of the neighboring lower-granularity level parameter in $lowMap$, we can take the replaced values with the highest weight to fill in the missing values (Algo. \ref{algo:replacePlow} $line_{1-5}$).

\noindent
\scalebox{0.65}{
\begin{minipage}{0.75\linewidth}
\begin{algorithm}[H]
$i_{replace, P_{CG}} \gets vWeightMap.top(1).key()$ \;
\If{$\not\exists p_{low}$}
    {
    $i_{m, P_{CG}} \gets i_{replace, P_{CG}}$ \;
    \For{$p_{cg} \in P_{CG}$}
        {
            \For{$w^{p_{cg}} \in Weak^H[p_{cg}]$}
                {
                \If{$i_{m, w^{p_{cg}}} = \o$}
                    {
                    $i_{m, w^{p_{cg}}} \gets \{i_{r, w^{p_{cg}}} \in I^D:i_{r,p_{cg}} = i_{m, p_{cg}}\}.getOne()$ \;
                    }
                }
        }
    }
\Else
    {
    $addMap(lowMap[i_{m, p_{low}}], i_{replace, P_{CG}},$ $vWeightMap[i_{replace, P_{CG}}])$ \;
    $addMap(lowMap, i_{m, p_{low}}, lowMap[i_{m, p_{low}}])$ \;
    }
\Return{$lowMap$}
\caption{$replaceNoPlow(D, H, lowMap,$ $vWeightMap, i_m, P_{CG}, p_{low}$)}
\label{algo:replaceNoPlow}
\end{algorithm}
\end{minipage}
}
\scalebox{0.65}{
\begin{minipage}{0.75\linewidth}
\begin{algorithm}[H]
\For{$i_{m, p_{low}} \in lowMap.keys()$}
    {
    $vWeightMap \gets lowMap[i_{m, p_{low}}].top(1)$ \;
    $i_{replace, p_{low}} \gets vWeightMap.key()$ \;
    \For{$i_m \in \{i_r \in I^D : i_{r,p_{low}} = i_{m, p_{low}}\}$}
        {
            $i_{m, P_{CG}} \gets i_{replace, P_{CG}}$ \;
            \For{$p_{cg} \in P_{CG}$}
                {
                    \For{$w^{p_{cg}} \in Weak^H[p_{cg}]$}
                        {
                        \If{$i_{m, w^{p_{cg}}} = \o$}
                            {
                            $i_{m, w^{p_{cg}}} \gets \{i_{r, w^{p_{cg}}} \in I^D:i_{r,p_{cg}} = i_{m, p_{cg}}\}.getOne()$ \;
                            }
                        }
                }
        }
    }
\caption{$replacePlow(lowMap, P_{CG}, H, D, p_{low})$}
\label{algo:replacePlow}
\end{algorithm}
\end{minipage}                              
}

\label{sec:expe}
\section{Experimental assessment}

\subsection{Technical environment and datasets}

To evaluate the 
effectiveness and efficiency of H-OLAPKNN, we implement our algorithms and conduct experiments with different datasets and compare it to other imputation methods. Our code is developed in Python 3.7 and is executed on a Intel(R) Core(TM) i5-10210U 1.60~GHz CPU with a 16~GB RAM. Data are integrated in R-OLAP format with Oracle 11g.
The distance metrics that we use are like described in section \ref{sec:distancemetric}, for the word embedding based distance, we use Google's pre-trained word2vec model \footnote{\url{https://drive.google.com/file/d/0B7XkCwpI5KDYNlNUTTlSS21pQmM/edit?resourcekey=0-wjGZdNAUop6WykTtMip30g}}. 

We employ 3 real world datasets. The first dataset is a regional sale dataset (\textbf{RegionalSales}) storing sales data for a company across US regions
. The second (\textbf{IBRD}) 
and the third (\textbf{MIGA}) ones 
are data of world bank which are respectively the International Bank for Reconstruction and Development balance sheet data and the Multilateral Investment Guarantee Agency issued projects data. For each one of the datasets, we create a DW for our experiment. The link of the dataset source and more information can be found in our github\footnote{\url{https://github.com/Implementation111/H-OLAPKNN/}}.

\subsection{Experimental methodology}
We apply different missing rates (1\%, 5\%, 10\%, 20\%, 30\%, 40\%) for each attribute except for the primary keys. To generate missing data for an attribute, 
we sort randomly all the instances and remove attribute values of the first certain percentage of instances. For the effectiveness, since we focus on the qualitative data, we apply the accuracy (number of correctly replaced values divided by number of missing values) as the metric instead of the root mean square error (RMSE) \cite{Miao18} which is widely used but is only suitable for quantitative data imputation. For the efficiency, we get the run time of each algorithm. We carry out 20 tests for each dataset and get the average accuracy and run time.

We compare our proposed method with some other methods in the literature as baseline. \textbf{H-OLAPKNN(MI)}: Since the mutual information is widely used in the KNN based data imputation \cite{GARCIALAENCINA20091483,pan2015}. In this baseline, we apply our proposed OLAPKNN algorithm by using the mutual information instead of the dependency degree as the hierarchy distance weight. \textbf{KNN} \cite{Domeniconi04}: This method use the basic KNN algorithm to generate the replaced values for missing data \textbf{Mode}: The Mode method simply replace the missing data with the most frequent non-empty value of the attribute in the table. 

\subsection{Results and analysis}

For each dataset, the optimal $k$ of KNN is different. So we test with different $k$ values between 1 and 20 and choose the best one for the experiment of each data set which are respectively 5, 4 and 8. For the $W_l$, 
we choose the weight with a better result as the weight for each dataset which are respectively $W_l^i$, $W_l^c$ and $W_l^i$. Then we compare the accuracy and the run time of each algorithm.

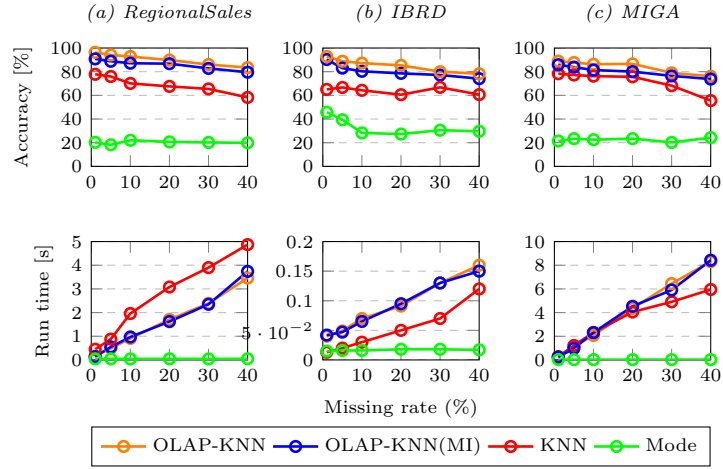
\begin{figure}
\centering
\begin{tikzpicture}
\begin{groupplot}[
   group style={
     group size=3 by 2,
     horizontal sep=1cm},
   width=0.9\linewidth/3,/tikz/font=\scriptsize]

\nextgroupplot[{title =\emph{(a) RegionalSales}},
    xmin=0, xmax=40,
    ymin=0, ymax=100,
    xtick={0,10,20,30,40,50},
    ytick={0,20,40,60,80,100},
    ylabel={Accuracy [\%]},
    legend pos=north west,
    ymajorgrids=true,
    grid style=dashed,
    legend style={at={(0,-0.9)},anchor=south west, legend columns=4}]
    \addplot[color=orange,mark=o, line width=1pt]
    coordinates {(1,96.3)(5,94.1)(10,92.8)(20,89.82)(30,86.03)(40,83.46)};
    \addplot[color=blue,mark=o, line width=1pt]
    coordinates {(1,90.8)(5,88.81)(10,87.13)(20,86.76)(30,82.77)(40,79.55)};
    \addplot[color=red,mark=o, line width=1pt]
    coordinates {(1,77.78)(5,75.93)(10,70.09)(20,67.58)(30,65.45)(40,58.26)};
    \addplot[color=green,mark=o, line width=1pt]
    coordinates {(1,20.32)(5,18.23)(10,22.08)(20,20.68)(30,20.23)(40,19.93)};
\nextgroupplot[{title =\emph{(b) IBRD}},  
    xmin=0, xmax=40,
    ymin=0, ymax=100,
    xtick={0,10,20,30,40,50},
    ytick={0,20,40,60,80,100},
    legend pos=north west,
    ymajorgrids=true,
    grid style=dashed,
    ]
\addplot[color=blue,mark=o, line width=1pt]
    coordinates {(1,90.3)(5,83.01)(10,80.33)(20,78.57)(30,77.2)(40,74.1)};
    \addplot[color=red,mark=o, line width=1pt]
    coordinates {(1,65)(5,66.67)(10,64.25)(20,60.56)(30,66.75)(40,60.656)};
    \addplot[color=orange,mark=o, line width=1pt]
    coordinates {(1,92.8)(5,88.81)(10,87.24)(20,85.34)(30,80)(40,78.25)};
    \addplot[color=green,mark=o, line width=1pt]
    coordinates {(1,45.75)(5,39.4)(10,28.28)(20,27.3)(30,30.58)(40,29.65)};

\nextgroupplot[{title =\emph{(c) MIGA}},
    xmin=0, xmax=40,
    ymin=0, ymax=100,
    xtick={0,10,20,30,40},
    ytick={0,20,40,60,80,100},
    legend pos=north west,
    ymajorgrids=true,
    grid style=dashed,]
\addplot[color=orange,mark=o, line width=1pt]
    coordinates {(1,88.89)(5,87.7)(10,86.27)(20,86.6)(30,79.02)(40,76.13)};
    \addplot[color=red,mark=o, line width=1pt]
    coordinates {(1,78.28)(5,77.19)(10,76.23)(20,75.66)(30,68.18)(40,55.56)};
    \addplot[color=blue,mark=o, line width=1pt]
    coordinates {(1,85.91)(5,83.83)(10,81.38)(20,80.07)(30,76.44)(40,73.76)};
    \addplot[color=green,mark=o, line width=1pt]
    coordinates {(1,21.45)(5,23.4)(10,22.58)(20,23.44)(30,20.23)(40,24.3)};
    
\nextgroupplot[
    xmin=0, xmax=40,
    ymin=0, ymax=5,
    xtick={0,10,20,30,40,50},
    ytick={0,1,2,3,4,5},
    ylabel={Run time [s]},
    legend pos=north west,
    ymajorgrids=true,
    grid style=dashed,
    legend style={at={(0,-0.9)},anchor=south west, legend columns=4}]
\addplot[color=orange,mark=o, line width=1pt]
    coordinates {(1,0.16)(5,0.49)(10,0.91)(20,1.71)(30,2.39)(40,3.46)};
       \addplot[color=blue,mark=o, line width=1pt]
    coordinates {(1,0.14)(5,0.55)(10,0.97)(20,1.62)(30,2.35)(40,3.74)};  
\addplot[color=red,mark=o, line width=1pt]
    coordinates {(1,0.45)(5,0.88)(10,1.96)(20,3.08)(30,3.9)(40,4.87)}; 
  
    \addplot[color=green,mark=o, line width=1pt]
    coordinates {(1,0.044)(5,0.047)(10,0.042)(20,0.048)(30,0.043)(40,0.049)};    
    \legend{OLAP-KNN, OLAP-KNN(MI),KNN,Mode}
\nextgroupplot[
    xlabel={Missing rate (\%)},   
    xmin=0, xmax=40,
    ymin=0, ymax=0.2,
    xtick={0,10,20,30,40,50},
    ytick={0,0.05,0.1,0.15,0.2},
    legend pos=north west,
    ymajorgrids=true,
    grid style=dashed,]
\addplot[color=orange,mark=o, line width=1pt]
    coordinates {(1,0.04)(5,0.049)(10,0.07)(20,0.091)(30, 0.13)(40,0.16)};
\addplot[color=red,mark=o, line width=1pt]
    coordinates {(1,0.013)(5,0.02)(10,0.03)(20,0.05)(30,0.07)(40,0.12)}; 
    \addplot[color=blue,mark=o, line width=1pt]
    coordinates {(1,0.042)(5,0.047)(10,0.065)(20,0.095)(30, 0.13)(40,0.15)};  
    \addplot[color=green,mark=o, line width=1pt]
    coordinates {(1,0.015)(5,0.016)(10,0.016)(20,0.018)(30,0.018)(40,0.017)};
\nextgroupplot[
    xmin=0, xmax=40,
    ymin=0, ymax=10,
    xtick={0,10,20,30,40},
    ytick={0,2,4,6,8,10},
    legend pos=north west,
    ymajorgrids=true,
    grid style=dashed,]
\addplot[color=orange,mark=o, line width=1pt]
    coordinates {(1,0.23)(5,1.17)(10,2.07)(20,4.36)(30,6.45)(40,8.3)};
\addplot[color=red,mark=o, line width=1pt]
    coordinates {(1,0.28)(5,1.22)(10,2.32)(20,4.05)(30,4.91)(40,5.95)}; 
    \addplot[color=blue,mark=o, line width=1pt]
    coordinates {(1,0.24)(5,0.98)(10,2.31)(20,4.53)(30,5.91)(40,8.42)};   
    \addplot[color=green,mark=o, line width=1pt]
    coordinates {(1,0.0262)(5,0.0265)(10,0.027)(20,0.027)(30,0.029)(40,0.028)};  
\end{groupplot}
\end{tikzpicture}
\caption{Results of experiments}
\label{result}
\end{figure}

\subsubsection{Accuracy}

The accuracy result is shown as Fig. \ref{result}. We can see that the proposed H-OLAPKNN algorithm outperforms all the other baseline algorithms for each dataset. The Mode method has the worst result since it is too simple and it takes nothing into account. Compared to the mutual information, we observe that using the dependency degree as the hierarchy distance weight can help us get a more accurate result as it considers the ordered dependency instead of the inter-dependency. Compared to the basic KNN method, the H-OLAPKNN returns a better accuracy results since it considers the structure of the DW and take the dependencies among the attributes into account. The accuracy of H-OLAPKNN, H-OLAPKNN(MI) and KNN decreases with the increase of the missing rate because when there are more missing data, 1) we have less complete data for getting more precise distance scores and 2) it is more likely that the proper replaced data do not exist in the table.

\subsubsection{Run time}
The run time result is shown as Fig \ref{result}. The Mode algorithm costs less time since it is the simplest method. The run time of the other three algorithms changes linearly with respect to the missing rate. There is no big difference between H-OLAPKNN and H-OLAPKNN(MI) since they are only different in terms of the calculation of hierarchy distance weight. The OLAPKNN costs less time than KNN for dataset \textbf{RegionalSales}, but more time for the other two datasets. This is because the hierarchical imputation complete most of the data of \textbf{RegionalSales} so that it takes less time for OLAPKNN to create the candidate list and compare the similarities.

\section{Conclusion and future work}

In this paper, we propose a DW dimensional data imputation method by combining hierarchical imputation with a novel KNN-based algorithm. We first define a 4-level distance calculation method for dimension instances by taking advantage of the DW dimension structure. Then by applying the proposed distances, we define the KNN-based algorithm. The advantage of the proposed algorithm is that it takes the dimension structure complexity into account and is able to make replaced values conform to the dependency constraints of the hierarchies. Our proposal is validated by a series of experiments and is proved to outperform other baselines in the literature. It increases the dimension data imputation accuracy by up to 25.2\% with respect to the basic KNN imputation.

In the future, we will extend our method for the imputation of numerical data in the dimensions and facts. We also intend to generalize the method for non-DW data by proposing an approach automatically modelling them in OLAP. 


\label{sec:conclu}

%
%
%
\bibliographystyle{splncs04}
\bibliography{mybibliography}

\end{document}